\UseRawInputEncoding

\documentclass[journal]{IEEEtran}
\usepackage{cite}
\ifCLASSINFOpdf
   \usepackage[pdftex]{graphicx}
   \graphicspath{{../pdf/}{../jpeg/}}
   \DeclareGraphicsExtensions{.pdf,.jpeg,.png}
\else
   \usepackage[dvips]{graphicx}
   \graphicspath{{../eps/}}
   \DeclareGraphicsExtensions{.eps}
\fi
\usepackage{amsmath}
\usepackage{algorithm}
\usepackage{algorithmic}
\usepackage{mathtools}
\usepackage{array}
  \usepackage[caption=false,font=footnotesize]{subfig}

\usepackage{stfloats}
\ifCLASSOPTIONcaptionsoff
  \usepackage[nomarkers]{endfloat}
 \let\MYoriglatexcaption\caption
 \renewcommand{\caption}[2][\relax]{\MYoriglatexcaption[#2]{#2}}
\fi

\usepackage{float}
\usepackage{stfloats}
\ifCLASSOPTIONcaptionsoff
  \usepackage[nomarkers]{endfloat}
 \let\MYoriglatexcaption\caption
 \renewcommand{\caption}[2][\relax]{\MYoriglatexcaption[#2]{#2}}
\fi
\usepackage{multicol}%,multienum

\usepackage{multirow}

\usepackage{url}
\usepackage{makecell}%表格中字体居中
\usepackage{color}%字体标色
\begin{document}
\title{Echo State Network based Symbol Detection in Chaotic Baseband Wireless Communication}
\author{Hui-Ping~Yin, Chao~Bai, and~Hai-Peng~Ren,~\IEEEmembership{Member,~IEEE}

\thanks{H.-P. Yin, H.-P. Ren are with the Shaanxi Key Laboratory
of Complex System Control and Intelligent Information Processing, Xi'an University of Technology, Xi'an, 710048 China (e-mail: huipingyin@qq.com, renhaipeng@xaut.edu.cn).}
\thanks{C. Bai is with the Xi'an Key Laboratory of Intelligent Equipment, Xi'an Technological University, Xi'an, 710021 China (e-mail: baic@xatu.edu.cn).}}

\maketitle

\begin{abstract}
In some Internet of Things (IoT) applications, multi-path propagation is a main constraint of communication channel. Recently, the chaotic baseband wireless communication system (CBWCS) is promising to eliminate the inter-symbol interference (ISI) caused by multipath propagation. However, the current technique is only capable of removing the partial effect of ISI, due to only past decoded bits are available for the suboptimal decoding threshold calculation. However, the future transmitting bits also contribute to the threshold. The unavailable future information bits needed by the optimal decoding threshold are an obstacle to further improve the bit error rate (BER) performance. Different from the previous method using echo state network (ESN) to predict one future information bit, the proposed method in this paper predicts the optimal threshold directly using ESN. The proposed ESN-based threshold prediction method simplifies the symbol decoding operation by removing the threshold calculation from the transmitting symbols and channel information, which achieves better BER performance as compared to the previous method. The reason for this superior result lies in two folds, first, the proposed ESN is capable of using more future symbols information conveyed by the ESN input to get more accurate threshold; second, the proposed method here does not need to estimate the channel information using Least Square method, which avoids the extra error caused by inaccurate channel information estimation. By this way, the calculation complexity is decreased as compared to the previous method. Simulation results and experiment based on a wireless open-access research platform under a practical wireless channel, show the effectiveness and superiority of the proposed method.
\end{abstract}


% \fi
\IEEEpeerreviewmaketitle

\section{Introduction}
\IEEEPARstart{M}{achine} learning (ML) has been recognized as the technology for solving dynamic system prediction problems recently \cite{Lecun2015Deep,bkassiny2012survey}. With the development in computing power and massive data collection and store ability, ML has been applied in more and more scientific research fields including wireless communication system for prediction, optimization, etc. \cite{sun2019application,jagannath2019machine,chen2019artificial,mosleh2017brain,8100870}. Reservoir computing (RC), as a branch of machine learning, is very suitable for short term memory and prediction\cite{lukovsevivcius2009reservoir,ButcherReservoir,manjunath2013echo,van2017advances}. RC has been proposed to estimate channel information \cite{liao2015channel,ZhaoEcho} and to predict the chaotic time series in chaos-based wireless communication system in order to improve the bit error rate (BER) performance \cite{jaeger2004harnessing,Ren2020performance}.

As a promising application, the efforts to use chaos in communication field started from early 1990s \cite{hayes1993communicating,rosa1997noise,corron1997new}, but most of initial researches were concentrated on the theoretical analysis, until ref \cite{argyris2005chaos} reported that chaos was successfully used in a commercial wired fiber-optic channel to achieve higher bit transmission rate in 2005. With this successful application of chaos as a landmark, the research on communication with chaos has moved towards practical communication channel \cite{ren2013wireless,corron2010matched,corron2015chaos,ren2014robustness,ren2016experimental,bai2016differential,ren2017secure,kaddoum2016wireless,bai2018chaos,bai2019digital,9098915}. The wireless channel is a complex practical channel, due to limited bandwidth, multipath propagation, Doppler shift, and complex noise, which cause serious distortion of the signal transmitted in it. Reference \cite{ren2013wireless} has proved the Lyapunov spectrum invariance property of chaotic signal after transmitting through wireless channels. At the same time, more properties of chaos promising to improve communication performance have been reported, including that chaotic signal has a very simple corresponding matched filter to maximize the signal to noise ratio \cite{corron2010matched,corron2015chaos}, and the chaos property can be used to resist inter-symbol interference (ISI) \cite{yao2017chaos,yao2019experimental}. As for the property of chaos, which can be used for eliminating ISI caused by multipath, it is Lyapunov invariance property \cite{ren2013wireless} that can be used in the symbol decoding process to elimate ISI. Since the ISI in wireless channel restricts the performance of chaotic baseband wireless communication system (CBWCS), more efforts have been made to relieve the multipath effect, including the ISI relief technique using suboptimal threshold given in \cite{yao2017chaos,yao2019experimental}. The symbol detection is critical in the CBWCS, because it directly determines if the promising performance of the system could be achieved. Theoretically, the inter-symbol interference (ISI) caused by multipath could be completely eliminated using the optimal decoding threshold \cite{yao2017chaos}. Practically, since the future symbols needed to calculate the optimal threshold are not available at the current time, the optimal threshold and the expected performance are hard to achieve. To improve the ISI relief performance of CBWCS, the critical challenge is how to get the optimal thresholds.

The unique feature of CBWCS is using a chaotic signal as the baseband signal. The chaotic signal is equivalently generated by an underlying deterministic dynamical system. To retrieval the information (pattern) conveyed by the signal, a variety of neural networks are potential options, such as feedforward neural network, recurrent neural network (RNN), and recently popular deep neural network (DNN) and convolutional neural network (CNN) etc. The most attracting feature of the feedforward neural network, DNN and CNN is complex function approximation ability, however, for the signal generated by a dynamical system, recurrent neural network is more suitable because of its unique dynamical storage capacity with respect to other candidates. But the shortcoming of RNN is its difficulty in training process, especially using the gradient-descent-based methods \cite{279181}. Fortunately, as one of reservoir computing method, ESN was proposed to drastically reduce the computational complexity by a linear regression method and facilitate the hardware integration \cite{lukovsevivcius2009reservoir}. Thus, ESN become our first choice to process the chaotic signal in CBWCS, either for waveform prediction or for the decoding threshold prediction, because all these operations involve the dynamical property of the dynamical system. To further clarify the reason for selecting ESN in this work, the features of several commonly used neural networks, are compared as shown in table I. ESNs have been used to model and predict chaotic systems in \cite{lu2018attractor,zimmermann2018observing,antonik2018using,lu2017reservoir}. A method based on ESN is proposed to predict one future symbol in CBWCS that is used to calculate a more accurate threshold for symbol detection \cite{Ren2020performance}. Compared with the suboptimal threshold in \cite{yao2017chaos}, this method achieves better result in BER performance, but it still requires the complex threshold calculation process. However, it is very hard to predict more future bits waveform using the method in \cite{Ren2020performance}, due to the noise and the accumulated error of the iterative prediction of ESN. To improve ISI resistance performance from this idea becomes very restrictive although the further future bits also contribute to the ISI resistance performance.
\newcommand{\tabincell}[2]{\begin{tabular}{@{}#1@{}}#2\end{tabular}}
\begin{table}[ht]
\renewcommand{\arraystretch}{1.7}
\caption{The features of several commonly used networks}
\label{table0}
\centering
\begin{tabular}{|c|c|}
\hline\hline
Feature & Methods\\
\hline
Higher prediction precision & \tabincell{c}{DNN \cite{J2015Deep}, CNN \cite{Krizhevsky2017ImageNet}\\RNN \cite{2006Short}, ESN \cite{jaeger2004harnessing}}   \\
\hline
\tabincell{c}{Dynamical memory fit \\for dynamical signal} & \tabincell{c}{RNN \cite{2006Short,lukovsevivcius2009reservoir}\\ ESN \cite{lukovsevivcius2009reservoir,manjunath2013echo}} \\
\hline
Lower complexity & ESN \cite{mosleh2017brain,lukovsevivcius2009reservoir}\\
\hline
High computation cost & \tabincell{c}{DNN \cite{2010Understanding}, CNN \cite{2014Convolutional}\\RNN \cite{lukovsevivcius2009reservoir,mosleh2017brain}}\\
\hline
\tabincell{c}{Easy to implement\\ in hardware} & ESN \cite{jaeger2004harnessing,mosleh2017brain}\\
\hline
\end{tabular}
\end{table}

To further improve the performance, a new method based on ESN is proposed in this paper. The trained ESN is used to predict the optimal threshold directly in the proposed method, instead of predicting future time series from which the future bit is decoded and then used to calculate the corresponding threshold as done in \cite{Ren2020performance}. The predicted thresholds can be used to decode symbols directly, which greatly simplifies the symbol detection procedure of CBWCS as compared to the method in \cite{Ren2020performance}. The predicted threshold here contains more future symbol information so as to further relieve the ISI and improve BER performance.

The contributions of this work are summarized as follows.

1) To avoid the difficulties and defects of predicting iteratively the matched filter output waveform, then, predicting one future symbol, finally calculating the decoding threshold using the predicted one future symbol, the past decoded four symbols and the identified channel information for decoding the current symbol in \cite{Ren2020performance}, ESN is proposed in this work for directly predicting the decoding threshold.

2) ESN structure and parameters for the threshold prediction are investigated, which shows that the number of internal neurons is the dominant factor in this application.

3) The controlled comparative investigation shows the superiority of the proposed method in the sense of low computation cost and low bit error rate as compared to the previous methods, including the method in \cite{Ren2020performance}, the conventional maximum likelihood method, and the conventional method with complicated MMSE equalization.

The merits of the proposed method include:

i) Only one prediction is needed for the proposed method to decode per symbol, rather than eight times iterative prediction for the matched filter output in the previous method in \cite{Ren2020performance}. which avoids the accumulate error caused by iteratively prediction.

ii) In decoding stage, the proposed method directly predicts the decoding threshold without eight times iterative prediction, the complicated threshold calculation using the past symbols, one predicted future symbol and the identified channel information as done in \cite{Ren2020performance}, which reduces the computation cost of the proposed method.

iii) The predicted threshold in the proposed method is capable of containing more future symbols information rather than only one (inaccurate) future symbol information as method in \cite{Ren2020performance}. This point contributes to lower BER of the proposed method.

The rest part of the paper is organized as follows. Section II gives a brief introduction to CBWCS. Section III introduces the proposed ESN-based threshold prediction in details. Section IV gives simulation results and shows better performance achieved by the proposed method. Section V gives experimental results and Section VI gives conclusions.


\section{Problem Formulation}
The CBWCS was proposed in \cite{yao2017chaos,yao2019experimental}, which is shown by Fig. 1.
\begin{figure}[ht]
  \centering
  \includegraphics[width=3.4in,height=0.7in]{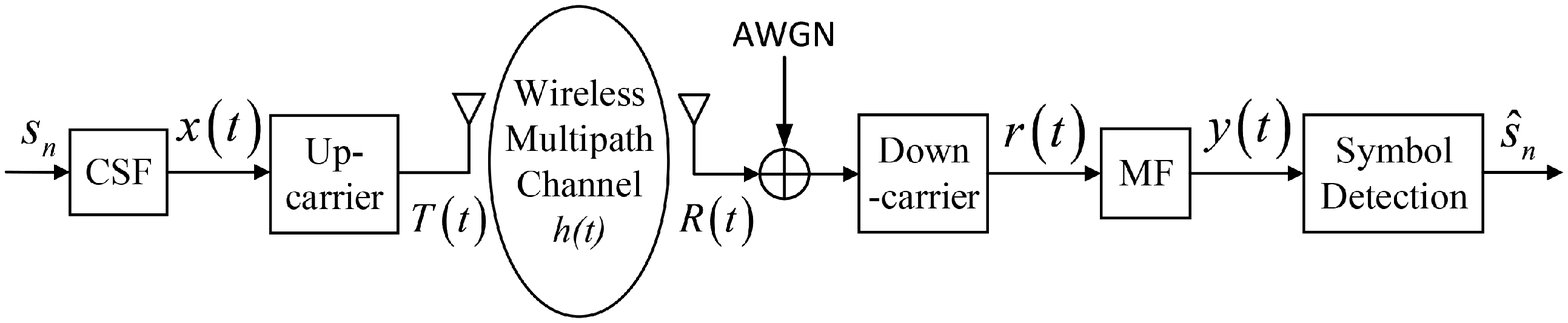}
  \caption{Block diagram of the CBWCS.}
\end{figure}

As shown in Fig. 1, the baseband signal $x\left( t \right) = \sum\limits_{ - \infty }^\infty  {{s_n} \cdot p\left( {t - {n \mathord{\left/ {\vphantom {n f}} \right. \kern-\nulldelimiterspace} f}} \right)} $ is generated by the chaotic shape forming filter (CSF)\cite{renchaotic} at the transmitter, where ${s_n} \in \left[ { - 1,1} \right]$ is the symbol to be transmitted and $p\left( t \right)$ is the basis function given by
\begin{equation}
%\begin{aligned}
 p\left( t \right)\! =\! \left\{ \begin{aligned}
&\left( {1\! -\! {e^{ - {\beta  \mathord{\left/
 {\vphantom {\beta  f}} \right.
 \kern-\nulldelimiterspace} f}}}} \right){e^{\beta t}}\left( {\cos \omega t \!- \!\frac{\beta }{\omega }\sin \omega t} \right),\left( {t < 0} \right)\\
&1\! - \!{e^{ - \beta \left( {t - {1 \mathord{\left/
 {\vphantom {1 f}} \right.
 \kern-\nulldelimiterspace} f}} \right)}}\left( {\cos \omega t\! -\! \frac{\beta }{\omega }\sin \omega t} \right),\left( {0 \le t < {1 \mathord{\left/
 {\vphantom {1 f}} \right.
 \kern-\nulldelimiterspace} f}} \right)\\
&0,\left( {t \ge {1 \mathord{\left/
 {\vphantom {1 f}} \right.
 \kern-\nulldelimiterspace} f}} \right)
\end{aligned} \right.,
%\end{aligned}
\end{equation}\\where $\omega $, $\beta $ are the parameters of the CSF satisfying $\omega  = 2\pi f$, and $f$ is the base frequency.

The baseband signal is chaotic, which is shown by the chaotic attractor in the phase plot and the positive Lyapunov Exponent (LE) \cite{yao2017chaos}. In this paper, LE is $ln2$. To be fit for channel transmission, $x\left( t \right)$ is modulated in the carrier frequency through up-carrier to derive the transmitted signal ready for antenna propagation, $T\left( t \right) = x\left( t \right)\cos \left( {2\pi {f_c}t} \right)$.

After the wireless channel transmission, the received signal, $R\left( t \right)$, passes through down-carrier to derive the received baseband signal given by $r\left( t \right) = R\left( t \right)\cos \left( {2\pi {f_c}t} \right)$, then a matched filter (MF) is used to maximize signal to noise ratio (SNR), $y\left( t \right) = g\left( t \right) * r\left( t \right)$, where the filter impulse response $g\left( t \right) = p\left( { - t} \right)$. The continuous signal $x\left( t \right)$ and $y\left( t \right)$ are sampled using over-sampling rate equal to 16 to get the sampled points, which are used for ESN training afterwards. Then the transmitted symbol sequences are decoded by sampling the filter output signal and comparing them with the corresponding decoding thresholds. However, the critical problem is how to obtain the optimal decoding threshold to eliminate the effect of ISI completely. According to the theoretical result in \cite{yao2017chaos}, the optimal threshold is

\begin{equation}
\theta  = {I_{past}} + {I_{future}},
\end{equation}
\noindent where ${I_{past}} = \sum\limits_{l = 0}^{L - 1} {\sum\limits_{i =  - \infty }^{ - 1} {{s_{n + i}}{I_{l,i}}} } $ represents the ISI caused by the past symbols and ${I_{future}} = \sum\limits_{l = 0}^{L - 1} {\sum\limits_{i = 1}^\infty  {{s_{n + i}}{I_{l,i}}} } $ represents the ISI caused by the future symbols,
${s_{n + i}}\left( {i < 0} \right)$ is the past transmitted symbols and ${s_{n + i}}\left( {i > 0} \right)$ is the future transmitting symbols. ${I_{l,i}}$ can be calculated by Eq. (3) (at the top of next page)
\begin{figure*}[ht]
%\normsize
%\begin{multicols}{2}
%\end{multicols}
\begin{small}
\begin{equation}
%\begin{aligned}
\mathclap{{I_{l,i}} \!=\! \left\{ \begin{array}{l}
{\alpha _l}{e^{ - \beta |{\tau _l} + \frac{i}{f}|}}\left( {2 -{e^{ - \frac{\beta }{f}}} - {e^{\frac{\beta }{f}}}} \right)\left( {A\cos \left( {\omega {\tau _l}} \right) + B\sin \left( {\omega {\tau _l}} \right)} \right),if\left( {|{\tau _l} + \frac{i}{f}| \ge \frac{1}{f}} \right)\\
{\alpha_l}\left\{{A\left({C\left({2\!-\!{e^{ - \frac{\beta }{f}}}} \right)\!- \!{C^{-1}}{e^{-\frac{\beta}{f}}}}\right)\cos \left({\omega {\tau_l}} \right)\! + \! B\left({C\left({2\!-\!{e^{-\frac{\beta }{f}}}}\right)\!+\!{C^{-1}}{e^{-\frac{\beta}{f}}}}\right)\sin \left({\omega{\tau_l}}\right)\!+\!1\!-\!|{\tau_l}f\!+\!i|}\right\},if\left({0\le|{\tau_l}\!+\!\frac{i}{f}| \!<\! \frac{1}{f}}\right),
\end{array}\right.\\}
%\end{aligned}
\end{equation}
\end{small}
\hrulefill
\vspace*{4pt}
\end{figure*}

\noindent where ${\alpha _l}$ and ${\tau _l}$ are the channel parameters (obtained by channel parameters estimation \cite{barhumi2003optimal}), $A \!=\! \frac{{\left( {{\omega ^2} - 3{\beta ^2}} \right)f}}{{4\beta \left( {{\omega ^2} + {\beta ^2}} \right)}}$, $B \!=\! \frac{{\left( {3{\omega ^2} - {\beta ^2}} \right)f}}{{4\omega \left( {{\omega ^2} + {\beta ^2}} \right)}}$ and $C\! =\! {e^{ - \beta |{\tau _l} + \frac{i}{f}|}}$.\\

From Eqs. (2) and (3), we know that the past information bits, the future information bits and the multipath information are needed to calculate the optimal threshold.

Due to the future symbols to be received are unknown at the current time, in \cite{yao2017chaos,yao2019experimental}, the suboptimal threshold calculated only from the past symbols was used for symbol detection. Although the suboptimal threshold achieved better performance as compared to that of the conventional non-chaotic baseband wireless communication system with conventional channel equalization method. It still leaves the space to improve the performance by considering the effect of the future symbols, along with this idea, reference \cite{Ren2020performance} proposed to use ESN to predict short-term future time series from the MF output and thus to predict one future symbol, by this way, the BER performance is improved as compared to the method in \cite{yao2017chaos,yao2019experimental}. However, the ESN in \cite{Ren2020performance} can only predict short-term future time series within the permitted precision, more future symbols prediction with high precision is restricted. To solve this problem, the ESN is employed to predict the optimal threshold directly in this work, it utilizes a massive short-term memory to build an accurate dynamic model and convert the waveform pattern into the more accurate threshold through the reservoir and the trained readout weights. In this method, the threshold calculation process is not needed, by this way, the computation cost is significantly decreased as compared to the method in \cite{Ren2020performance}, at the same time, because the past time series used in memory contains more than one bits future information, the proposed method achieves better performance than the previous method \cite{Ren2020performance}.

\section{ESN-based Threshold Prediction}
In the previous work, ESN is trained and used to predict the matched filter output signal at the receiver of CBWCS. The ESN used in \cite{Ren2020performance} is configured as one input scalar data (one sampling or iteratively predicted data), one output scalar data (corresponding to one predicted sampling data) and 80 internal neurons, after the training process, the ESN is used to predict 8 steps future output recursively. After predicting the first sampling point of the next symbol, we have to predict next sampling by using the first prediction, which is not actual waveform sampling (therefore, includes accumulate error if it is used iteratively), to update the reservoir states. Eight iterative predictions are needed, because, theoretically, only the eighth sampling point has the maximum signal to noise ratio for decoding the next symbol. Then the 8th prediction output is compared with zero, to obtain one future bit. After this step, the one future bit, together with the past decoded bits and the identified channel information, is used to calculate the decoding threshold $\theta  = {I_{fut1}} + {I_{past}}$.

\subsection{ESN-based Threshold Prediction Scheme}
In this paper, different from the method in \cite{Ren2020performance}, an ESN with $K$ input data and one-dimensional output is employed. The structure of the proposed ESN is given in Fig. 2.
\begin{figure}[ht]
  \centering
  \includegraphics[width=2.6in,height=1.1in]{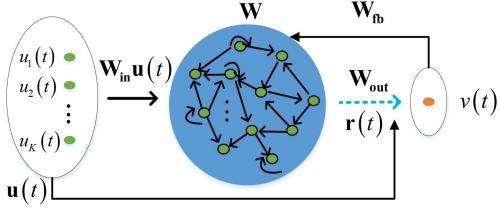}
  \caption{Structure of ESN. Solid bold arrows represent fixed synaptic connections, dotted arrows represent trainable connections.}
\end{figure}

In the input layer, the input vector is ${\bf{u}}(t) = {\left( {{u_1}(t),...,{u_K}(t)} \right)^T} \in {C^{K \times 1}}$, where $K$ represents the number of input units; the state vector of the reservoir is expressed as ${\bf{r}}(t) = {\left( {{r_1}(t),...,{r_N}(t)} \right)^T} \in {C^{N \times 1}}$, where $N$ is the number of internal neurons in the reservoir; one dimensional output is $v(t)$. The weighted input ${{\bf{W}}_{{\bf{in}}}}{\bf{u}}\left( t \right)$ and the weighted feedback
${{\bf{W}}_{{\bf{fb}}}}{v}\left( {t - 1} \right)$ are combined with the previous reservoir state ${\bf{Wr}}\left( {t - 1} \right)$ to produce current reservoir state ${\bf{r}}\left( t \right)$. Then the ${\bf{r}}\left( t \right)$ and input ${\bf{u}}\left( t \right)$ are combined through the output weights ${{\bf{W}}_{{\bf{out}}}} \in {C^{1 \times \left( {K + N} \right)}}$ to obtain the output $v(t)$. In this ESN, only the output weights ${{\bf{W}}_{{\bf{out}}}}$ need to be trained in the training phase, where the input weight matrix ${{\bf{W}}_{{\bf{in}}}} \in {C^{N \times K}}$, the internal weight matrix ${\bf{W}} \in {C^{N \times N}}$, and the feedback weight matrix ${{\bf{W}}_{{\bf{fb}}}} \in {C^{N \times 1}}$ are randomly chosen constants.

In this paper, the ESN is proposed to predict the optimal decoding threshold for symbol detection at the receiver, and it acts as a black box for modeling the matched filter output $y\left( t \right)$ of a symbol to the corresponding optimal decoding threshold $\theta $, then used to predict the optimal decoding threshold, $\hat \theta $. In CBWCS, the matched filter output is a continuous signal, which is sampled at the over-sampling rate ${N_s}$ (here  ${N_s} = 16$). The block diagram of the ESN-based threshold prediction and symbol decoding is shown in Fig. 3.
\begin{figure}[ht]
  \centering
  \includegraphics[width=3.3in,height=1.2in]{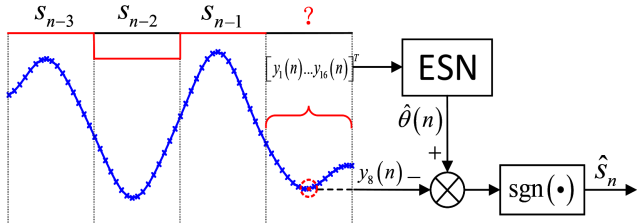}
  \caption{Block diagram of ESN-based threshold prediction in CBWCS.}
\end{figure}

As seen from Fig. 3, the received matched filter output corresponding to the current symbol ${s_n}$ at the receiver is sampled at the over-sampling rate, by this way, the sampled data recorded as $\left[ \begin{array}{l}
{y_1}\left( n\right)\\
 \vdots \\
{y_{16}}\left( n\right)
\end{array} \right]$, is used as the input of the ESN to calculate the output of ESN, which is the prediction of the optimal decoding threshold, i.e., $\hat \theta \left( n \right)$. The underlying principle to use such configuration lies in the assumption that the past and the current waveform contain the information of the future symbol and its corresponding waveform. Therefore, we do not predict the future waveform, instead, we directly predict the threshold using the information stored in the trained ESN and the mapping ability of ESN, by this way, avoid the complicated calculation for the threshold contributed by the past symbols and that by the future symbol in Eq. (2), thus the proposed method simplifies the decoding process.

However, simplification is achieved at the cost of the increased number of the ESN input as well as the increased complexity in the training process. But we confirm this cost is deserved by the results afterwards. The reasons are given in the following, the first reason is that the method in this work avoid the accumulated error by using the iterative prediction in \cite{Ren2020performance}. The second reason is that the proposed method in this work significantly decreases the computational complexity by avoiding the threshold computation using the channel information, the past symbols and the future symbol for every symbol decoding, also by predicting one symbol decoding threshold per time rather than eight times for one future symbol prediction in method \cite{Ren2020performance}. The third reason is the ESN trained in this work contains more future symbols information, rather than just one future symbol information being used in \cite{Ren2020performance}, which leads to a significant BER performance improvement.

\subsection{Training Data}
From the feature of the CSF used in the CBWCS and Eq. (2), we know that, the farther the bit locates with respect to the current bit to decode, the smaller effect on the decoding of the current bit. Therefore, to predict the optimal decoding threshold of the current symbol, the chaotic signal corresponding to the past symbols, current symbol and future symbols should be remembered in the ESN.
%\begin{figure}[ht]
%  \centering
%  % Requires \usepackage{graphicx}
%  \includegraphics[width=2.8in,height=1.8in]{Fig4.eps}
%  \caption{The basis function $p\left( t \right)$ for $\beta  = \ln 2$ and $f = 1$.}
%\end{figure}

In CBWCS, the current symbol waveform is mainly affected by 4 future symbols and 4 past symbols \cite{Ren2020performance}. However, the training process in our method is on-line performed, more bits are considered, more training data is required, it will cost too much time for training. Thus, the sampling points corresponding to 3 past symbols, one current symbol and 3 future symbols are used for training the ESN, which reduces the number of the training data needed and training time with not too much performance deterioration.

3 past symbols, 1 current symbol and 3 future symbols, totally 7 symbols are considered. There are ${2^7}$ possible combinations of these 7 symbols, therefore, ${2^7} \times 7 = 896$ training symbols should be considered during the training process. Considering the effect of the initial condition (initial state of inner neurons) of the ESN, 100 random symbols are inserted in the front of the aforementioned training data to wash out the initial memory in the dynamic reservoir. By this way, 996 symbols data, recorded as $\left( {\begin{array}{*{20}{c}}
{{y_1}\left( 1 \right)}& \ldots &{{y_1}\left( {{\rm{996}}} \right)}\\
 \vdots & \ddots & \vdots \\{{y_{16}}\left( 1 \right)}& \cdots &{{y_{16}}\left( {{\rm{996}}} \right)}
\end{array}} \right)$, are used as inputs to train ${{\bf{W}}_{{\bf{out}}}}$. The corresponding calculated optimal decoding thresholds, i.e, $\theta \left( 1 \right), \cdots ,\theta \left( {996} \right)$, are the teacher outputs of the ESN.

\subsection{Training of the Echo State Network}
Given the ESN structure and the input-output sequences, the output weights ${{\bf{W}}_{{\bf{out}}}}$ can be trained. The available input-output sequences are divided into two parts:

1) the initial part, which is used to remove the effect of initial condition of the ESN;

2) the training part, which is used to train the output weights ${{\bf{W}}_{{\bf{out}}}}$.

The output weights are determined using the teacher outputs and inputs. Then the trained output weights can be used in the prediction task without any further modifications, which guarantees the low computation requirement of the ESN application. The training process is summarized in the following.

{\bfseries{\emph{Step}}} 1: ${{\bf{W}}_{{\bf{in}}}}$, ${{\bf{W}}_1}$ and ${{\bf{W}}_{{\bf{fb}}}}$ are randomly generated as a uniform distribution in [-1, 1]. ${{\bf{W}}_{{\bf{in}}}}$, ${{\bf{W}}_{{\bf{fb}}}}$ will not change during the training process, and ${\bf{W}}$ can be obtained by rescaling ${{\bf{W}}_1}$ through ${\bf{W}} = \left( {{{\rho \left( {\bf{W}} \right)} \mathord{\left/
 {\vphantom {{\rho \left( {\bf{W}} \right)} {|{\lambda _{\max }}|}}} \right.
 \kern-\nulldelimiterspace} {|{\lambda _{\max }}|}}} \right){{\bf{W}}_1}$, where $|{\lambda _{\max }}| < 1$ is the largest eigenvalue of ${{\bf{W}}_1}$ and $\rho \left( {\bf{W}} \right)$ is the spectral radius of ${\bf{W}}$.

{\bfseries{\emph{Step}}} 2: The network's internal state vector ${\bf{r}}$ and output threshold $\theta $  are initialized to zero, i.e., ${\bf{r}}\left( 0 \right) = {\bf{0}}$, $\theta \left( 0 \right) = 0$. Feed the training data into the reservoir. The new state of reservoir is updated by
\begin{equation}
{\bf{r}}\left( {n+1} \right) = f\left( {{{\bf{W}}_{{\bf{in}}}}{\bf{u}}\left( {n+1} \right) + {\bf{Wr}}\left( n \right) + {{\bf{W}}_{{\bf{fb}}}}\theta \left( n \right)} \right),
\end{equation}

\noindent where $f$ is the tanh function, ${\bf{u}} = {\left[ {{y_1}, \cdots ,{y_{16}}} \right]^T}$ is the input vector, here, $n = 0, \cdots, {n_{\max }}-1$, and ${n_{\max }} = 996$ is the number of training data. After the update, ${\bf{r}}\left( 1 \right), \cdots ,{\bf{r}}\left( {{n_{\max }}} \right)$ are obtained.

{\bfseries{\emph{Step}}} 3: Wash out the initial memory in the dynamic reservoir. To avoid the impact of the different initial values of ${\bf{r}}$ on the output weight matrix ${{\bf{W}}_{{\bf{out}}}}$, the data for $n = 1,...,{n_0}\left( {{n_0} = 100} \right)$, is not used for ${{\bf{W}}_{{\bf{out}}}}$ calculation. For each $n = {n_0} + 1,...,{n_{\max }}$, collect the input data ${\bf{u}}\left( n \right)$ and the corresponding network state ${\bf{r}}\left( n \right)$ as rows into a state and input collecting matrix ${\bf{R}} = \left( {\begin{array}{*{20}{c}}{{\bf{r}}\left( {{n_0} + 1} \right)}& \ldots &{{\bf{r}}\left( {{n_{\max }}} \right)}\\{}&{}&{}\\{{\bf{u}}\left( {{n_0} + 1} \right)}& \cdots &{{\bf{u}}\left( {{n_{\max }}} \right)}\end{array}} \right)$, which is the size of $\left( {N + K} \right) \times \left( {{n_{\max }} - {n_0}} \right)$. The corresponding desired output $\theta \left( n \right)$ is collected in ${\bf{T}} = \left[ {\theta \left( {{n_0} + 1} \right), \cdots ,\theta \left( {{n_{\max }}} \right)} \right]$ as described aforementioned.

{\bfseries{\emph{Step}}} 4: The output weights matrix ${{\bf{W}}_{{\bf{out}}}}$ is calculated by
\begin{equation}
{{\bf{W}}_{{\bf{out}}}} = {\bf{T}}{{\bf{R}}^T}{\left( {{\bf{R}}{{\bf{R}}^T} + {\lambda _r}{\bf{I}}} \right)^{ - 1}},
\end{equation}

\noindent where ${\bf{I}}$ is the identity matrix with proper dimension and   ${\lambda _r}$ is the regularization coefficient, typically, the value of   ${\lambda _r}$ is less than 1.

\subsection{Threshold Prediction}
After the ${{\bf{W}}_{{\bf{out}}}}$ is trained, the ESN with fixed ${{\bf{W}}_{{\bf{in}}}}$, ${\bf{W}}$, ${{\bf{W}}_{{\bf{fb}}}}$ and trained ${{\bf{W}}_{{\bf{out}}}}$ is ready for use. Assuming the neurons' states in the prediction phase start from ${n_{\max }}$, and $n = {n_{\max }} + 1$, where ${n_{\max }}$ is the size of the training data. The equations for the prediction phase can be given as
\begin{equation}
{\bf{r}}\left( n \right) = f\left( {{{\bf{W}}_{{\bf{in}}}}{\bf{u}}\left( n \right) + {\bf{Wr}}\left( {n-1}\right) + {{\bf{W}}_{{\bf{fb}}}} \theta \left( {n-1} \right)} \right),
\end{equation}

\begin{equation}
\hat \theta \left( n \right) = {{\bf{W}}_{{\bf{out}}}}\left[ {{\bf{r}}\left( n \right);{\bf{u}}\left( n \right)} \right],
\end{equation}

\noindent where the reservoir activation function $f$ in Eq. (6) is tanh function, and the new state vector ${\bf{r}}\left( n\right)$ is updated. In Eq. (7), $\hat \theta \left( n\right)$ is the prediction value of the corresponding optimal threshold through the reservoir state ${\bf{r}}\left( n \right)$ and the input vector ${\bf{u}}\left( n \right)$ for symbol ${s_n}$.

After the predicted decoding threshold $\hat \theta \left( n \right)$ is obtained, the symbol ${s_n}$ can be decoded according to the following equation given by
\begin{equation}
{\hat s_n} = {\mathop{\rm sgn}} \left( {{y_{\max }}\left( n \right) - \hat \theta \left( n \right)} \right),
\end{equation}

\noindent where ${y_{\max }}\left( n \right)$  is the maximum signal to noise ratio (SNR) point of the sampling data corresponding to symbol ${s_n}$, in this work, it is ${y_8}\left( n \right)$ as shown in Fig. 3.

\subsection{ESN Parameters Selection}
The performance of the network depends on the parameters to some extent. The parameters of ESN include structure parameters and statistical parameters. The structure parameters include the input number, the output number and reservoir size. The statistical parameters include the spectral radius $\rho $, the sparse degree SD, the initial value of ${{\bf{W}}_{{\bf{in}}}}$, ${{\bf{W}}_{{\bf{fb}}}}$, ${\bf{W}}$. The input and output number depends on the problem to be solved. The initial values of ${{\bf{W}}_{{\bf{in}}}}$, ${{\bf{W}}_{{\bf{fb}}}}$, ${\bf{W}}$ satisfy the statical constraints, including the spectral radius $\rho \left( {\bf{W}} \right)$ and the largest singular value of ${\bf{W}}$ are both less than one. ${{\bf{W}}_{{\bf{in}}}}$, ${{\bf{W}}_{{\bf{fb}}}}$ have little effect on the decoding BER performance. In the following, we address three important parameters of ESN and discuss their selection rules.

\subsubsection{Reservoir Size}
The reservoir size is one of the most important structure parameters. It is the number of neurons in the reservoir, in general, it is expected to achieve better performance by using larger reservoir size (RS), i.e., more neurons in the reservoir. However, as the number of the neurons increasing, the training complexity and online predicting complexity increase too. Since 3 past symbols, one current symbol and 3 future symbols are considered for decoding the current symbol in this work, the states of 7 symbols need to be recorded in ESN. The corresponding sampling points are $7 \times {N_s} = 112$, where the over-sampling rate ${N_s} = 16$. Based on the above consideration, the number of neurons in the reservoir is set as 112.

\subsubsection{Spectral Radius}
The spectral radius is a statistical parameter of ESN. The closer to 1 the spectral radius is, the slower the decay of the network response to the input. Since the decoding threshold of the current symbol is related to the past and future symbols, a substantial short-term memory is required for our threshold prediction task, thus the spectral radius in our task should be set relatively high. According to the prediction performance, different values of spectral radius are tested on the training data manually, so that the spectral radius is set as 0.9 in this work for the best performance.

\subsubsection{Sparse Degree}
The sparse degree (SD) represents the ratio of nonzero elements to total elements in the internal weights ${\bf{W}}$. To achieve the echo state property, the network needs to be sparsely and randomly connected. The value of SD is generally set from 1$\%$ to 5$\%$. In our method, numerous tests demonstrate that the value of SD has little effect on the decoding BER performance. We set it as 0.02 in this work.

A more systematic cross-optimization of the parameters would probably improve the training result. However, it leaves little margin for improvement, which is not the focus of this paper.
\section{Simulation Result}
In this section, simulations are performed to test the BER performance of the ESN-based threshold prediction method. In CSF, $f = 1$, $\beta  = \ln 2$, and the parameters of ESN used are shown in table II.

\begin{table}[ht]
\renewcommand{\arraystretch}{1.5}
  %\centering
  \caption{Reservoir parameters used for threshold prediction}\label{table_example}
  \centering
  \begin{tabular}{c|c}
  \hline\hline%\noalign{\smallskip}
  Parameter&Value\\
  \hline%\noalign{\smallskip}\noalign{\smallskip}
  N&112\\
  SD&0.02\\
  $\rho $ &0.9\\
  \hline%\noalign{\smallskip}
  \end{tabular}
  \end{table}

We simulated the single path and multipath conditions using the decoding threshold predicted by ESN. The BER comparison of different methods is performed, which includes the methods using different thresholds, i.e., zero threshold, the suboptimal threshold proposed in \cite{yao2017chaos,yao2019experimental}, the improved threshold using the past bits and one predicted future bit in \cite{Ren2020performance}, and the proposed predicted optimal threshold in this paper; and the method using Minimum Mean Square Error (MMSE) algorithm for channel equalization.

\subsection{Simulation under Single Path}
The BER comparison results under single path using thresholds $\theta  = 0$, $\theta  = {I_{past}}$ in \cite{yao2017chaos}, $\theta  = {I_{fut1}} + {I_{past}}$ in \cite{Ren2020performance}, $\theta  = \hat \theta $, and the conventional method with MMSE equalization in this paper are plotted in Fig. 4.
\begin{figure}[ht]
  \centering
  \includegraphics[width=3.0in,height=2.2in]{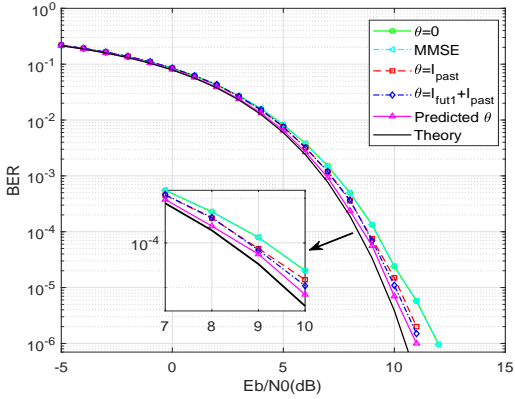}
  \caption{ BER results under single path. Black solid line is the theory BER curve for single path case, magenta solid line with upper triangular markers is the BER curve obtained using the threshold proposed in this paper, blue dash-dotted line with diamond markers is the BER curve obtained using the threshold in \cite{Ren2020performance}, red dashed line with square markers is the BER curve obtained using the threshold in \cite{yao2017chaos,yao2019experimental}, cyan dash-dotted line with upper triangular markers is the BER curve obtained using the zero threshold with MMSE equalization, and green solid line with circle markers is the BER curve obtained using zero threshold.}
\end{figure}

From Fig. 4, we can see that the predicted decoding threshold leads to the best result as compared to the other methods although the gap between methods are not large for single path case.

\subsection{Simulation under Multipath Condition}
The simulation results of different methods under 2 and 3 paths are shown in Figs. 5 and 6 respectively. The channel fading $h\left( {{\tau _l}} \right) = {e^{ - \gamma {\tau _l}}}$, where $\gamma $ is damping coefficient \cite{dottling2009radio}, and $L$ is the number of multipaths. In the following simulation, $\gamma  = 0.6$.
%parameters $\gamma  = 0.6$, ${\alpha _0} = 1$, ${\tau _1} = 1$, ${\alpha _1} = {e^{ - \gamma {\tau _1}}}$ for $L = 2$, ${\alpha _0} = 1$, ${\tau _1} = 1$, ${\alpha _1} = {e^{ - \gamma {\tau _1}}}$, ${\tau _2} = 2$, ${\alpha _2} = {e^{ - r{\tau _2}}}$ for $L = 3$ are used for test.
\begin{figure}[ht]
  \centering
  \includegraphics[width=2.8in,height=2.2in]{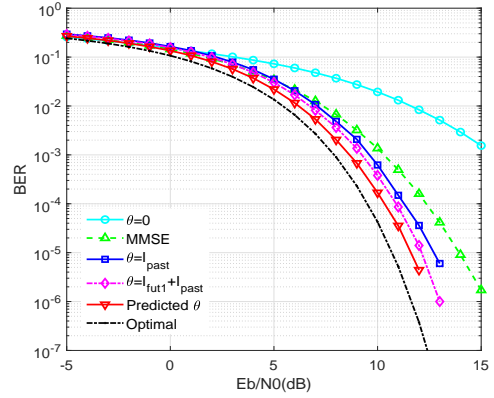}
  \caption{BER results of different methods under 2 paths. Black dash-dotted line is the optimal BER corresponding to $L = 2$, red solid line with lower triangular markers is the BER curve using the predicted threshold proposed in this paper, magenta dash-dotted line with diamond markers is the BER curve using the threshold in \cite{Ren2020performance}, blue solid line with square markers represents the BER curve using the threshold in \cite{yao2017chaos,yao2019experimental}, green dashed line with upper triangular markers is the BER curve using zero threshold with MMSE equalization, and the cyan solid line with circle markers is the BER curve using zero threshold.}
\end{figure}

\begin{figure}[ht]
  \centering
  \includegraphics[width=2.8in,height=2.2in]{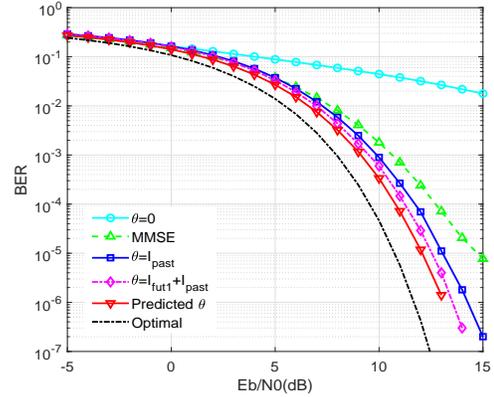}
  \caption{BER results of different methods under 3 paths. Black dash-dotted line is the optimal BER corresponding to $L = 3$, red solid line with lower triangular markers is the BER curve using the predicted threshold proposed in this paper, magenta dash-dotted line with diamond markers is the BER curve using the threshold in \cite{Ren2020performance}, blue solid line with square markers represents the BER curve using the threshold in \cite{yao2017chaos,yao2019experimental}, green dashed line with upper triangular markers is the BER curve using zero threshold with MMSE equalization, and the cyan solid line with circle markers is the BER curve using zero threshold.}
\end{figure}

The results using different methods for time-invariant 2 and 3 paths in Figs. 5 and 6 show that the CBWCS has the worst BER by using the decoding threshold without considering ISI from past and future symbols, i.e., $\theta  = 0$. Using the decoding threshold in \cite{yao2017chaos,yao2019experimental} and the threshold in \cite{Ren2020performance} achieve better BER performance as compared to the corresponding case using $\theta  = 0$. The BER using the threshold predicted by ESN in this paper is also superior to the conventional method with MMSE equalization, which is the lowest among all different methods. We can also see that the zero threshold leads to the worst BER performance, which is even worse as compared to the single path case. From Fig. 5, we can see an improvement of more than 0.5 dB using the proposed threshold with respect to the threshold in \cite{Ren2020performance} and more than 1 dB with respect to the suboptimal threshold proposed in \cite{yao2017chaos,yao2019experimental} when BER $ ={10^{ - 4}}$ for two paths case. For three paths case, the similar conclusion can be drawn as well.

In the above analysis, the wireless channel is time-invariant during the information transmission and the corresponding channel parameters are known for symbol decoding. However, the wireless is time-varying and the channel parameters are unknown in practice. In such case, we assume that the channel parameter is unchanged within one data frame, but is variant from one frame to the other. Here, the channel parameter, $\gamma$ obeys the uniform distribution in the range of [0.3, 0.9]. Figures 7 and 8 give the BER comparison results under a time-varying channel corresponding to $L = 2$ and $L = 3$.
\begin{figure}[ht]
  \centering
  \includegraphics[width=2.8in,height=2.2in]{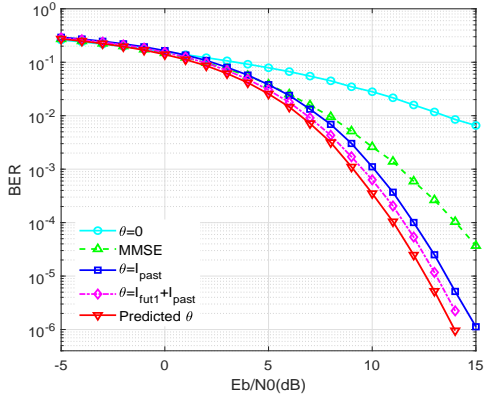}
  \caption{ BER results with different thresholds under 2 paths in a time-varying wireless channel. Red solid line with lower triangular markers is the BER curve using the predicted threshold proposed in this paper, magenta dash-dotted line with diamond markers is the BER curve using the threshold in \cite{Ren2020performance}, blue solid line with square markers represents the BER curve using the threshold in \cite{yao2017chaos,yao2019experimental}, green dashed line with upper triangular markers is the BER curve using zero threshold with MMSE equalization, and the cyan solid line with circle markers is the BER curve using zero threshold.}
\end{figure}

\begin{figure}[ht]
  \centering
  \includegraphics[width=2.8in,height=2.2in]{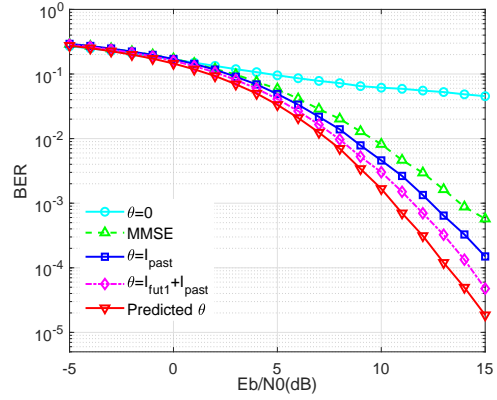}
  \caption{ BER results with different thresholds under 3 paths in a time-varying wireless channel. Red solid line with lower triangular markers is the BER curve using the predicted threshold proposed in this paper, magenta dash-dotted line with diamond markers is the BER curve using the threshold in \cite{Ren2020performance}, blue solid line with square markers represents the BER curve using the threshold in \cite{yao2017chaos,yao2019experimental}, green dashed line with upper triangular markers is the BER curve using zero threshold with MMSE equalization, and the cyan solid line with circle markers is the BER curve using zero threshold.}
\end{figure}

In our simulation, there are 2048 bits in one frame, which contains 996 training bits and 1052 information bits. The training bits are used for channel parameters estimation using the least square (LS) algorithm \cite{barhumi2003optimal} and ${{\bf{W}}_{{\bf{out}}}}$ training of ESN. The simulation results are obtained by averaging over 2000 frames. Due to the imperfect channel estimation, the simulation BERs are worse than the corresponding results in Figs. 5 and 6. However, for both two and three-path channels, the BER performance of the CBWCS using the predicted threshold of our method is better than the methods in \cite{Ren2020performance,yao2017chaos}, the method using zero threshold, the conventional method with MMSE equalization.

\section{Experimental Results}
To test the effectiveness of the proposed method in this paper, the wireless open-access research platform version 3 (WARP V3) is used in the experiment, which is designed by Rice University. The Xilinx Virtex6 LX240T FPGA is used for digital signal processing, two MAX2829 RF chips are for dual-channel, and 2.4GHz/5GHz dual-band transceiver, the maximum transmission power is 20dBm. To provide two ADC channels with sampling rates of 100 MSPS and two DAC channels with sample rates of 170 MSPS, the 12-bit low power analog/digital converter AD9963 is used, and two 10/100/1000 Ethernet interfaces (Marvell 88E1121R) are used for high-speed digital signal exchange with the Personal Computer (PC). The photo of the hardware system is shown in Fig. 9.
\begin{figure}[ht]
  \centering
  \includegraphics[width=2.6in,height=1.4in]{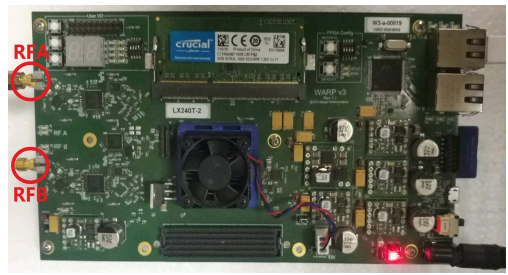}
  \caption{The WARP node photo}
\end{figure}

From Fig. 9, we can see that each WARP node has two radio antennas, which are referred to as RFA and RFB. Only RFA, operating at the 5GHz carrier frequency with 20MHz bandwidth, is used in our test. The information is transmitted frame by frame in our experiment, there are 2048 bits in one frame, which contains 996 frame header bits for ESN training and synchronization, 1052 bits as valid data. It is worth pointing out that the length of the frame can be changed with the case needed, the configuration in this paper is just a case for study.

\subsection{Single Path Channel Test}

At first, the test in the laboratory is carried out, the test scenario is shown as Fig. 10(a), and the identified channel parameters are shown in Fig. 10(b).
\begin{figure*}[ht]
\centering
\subfloat[]{\includegraphics[width=2.2in,height=1.5in]{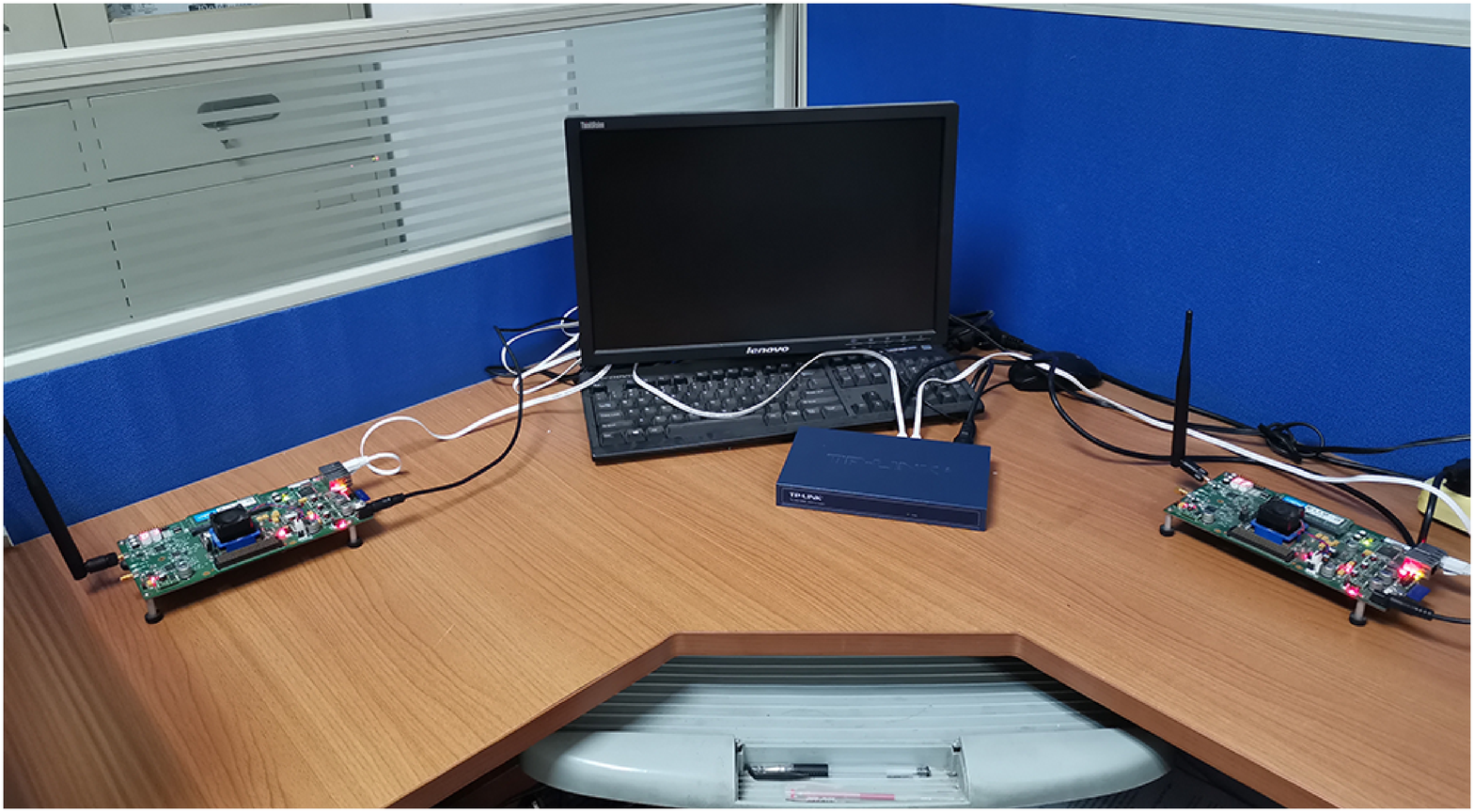}%
\label{first_case}}
\hfil
\subfloat[]{\includegraphics[width=2.3in,height=1.6in]{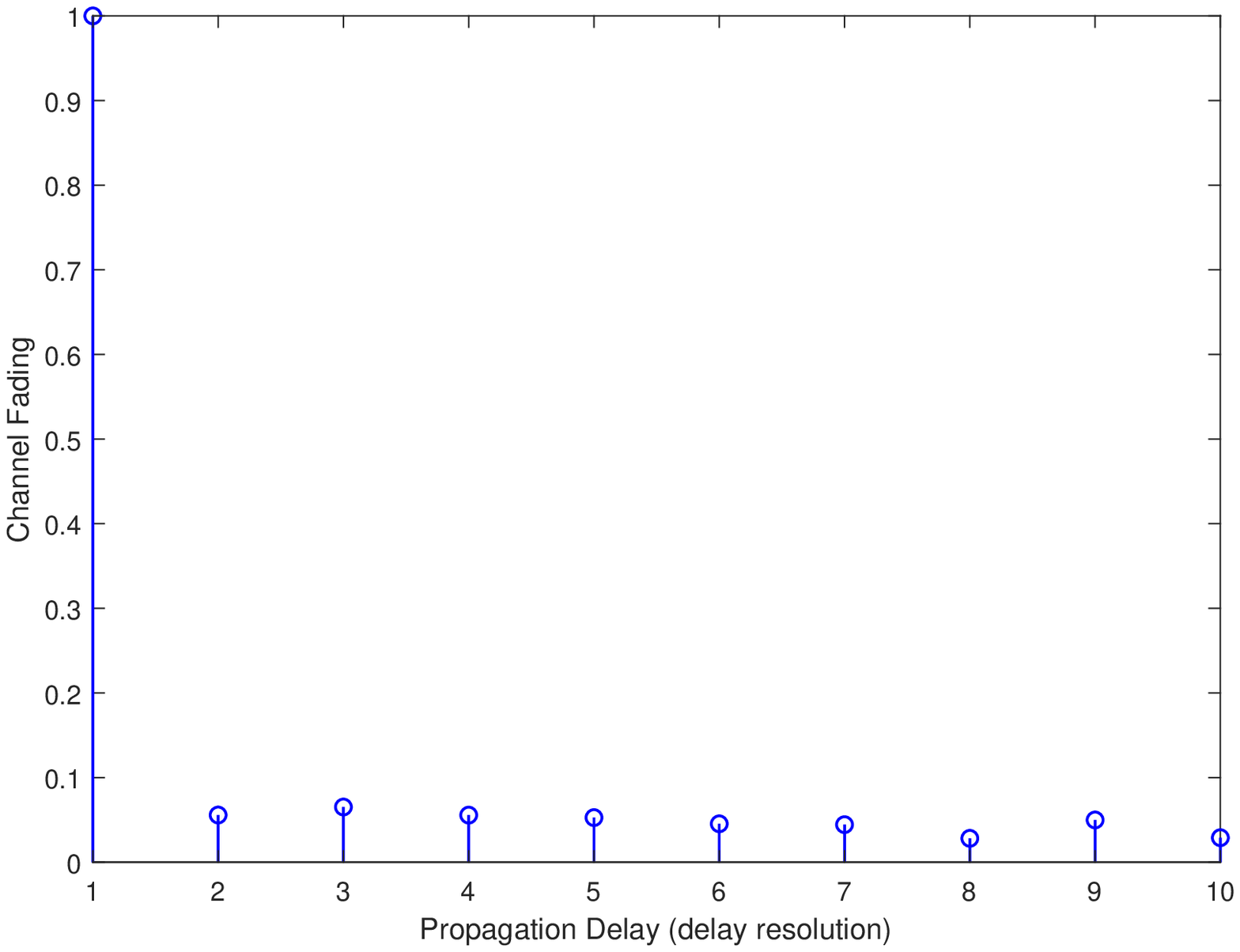}%
\label{thirdcase}}
\caption{(a) The photo of the test scenario, (b) the estimated channel parameters.}
\label{figsim}
\end{figure*}

From Fig. 10, We can see that the path with channel fading 1 is the main path, the channel fading of other paths is weak. The unit of the x-axis in Fig. 10(b) is the delay resolution, which is 0.25$\mu s$ in our system.

The BER comparison result under single path using different decoding methods is shown by Fig. 11, in which the horizontal axis is the transmission power, which is used to simulate the variation of SNR.
\begin{figure}[ht]
  \centering
  \includegraphics[width=2.8in,height=2.2in]{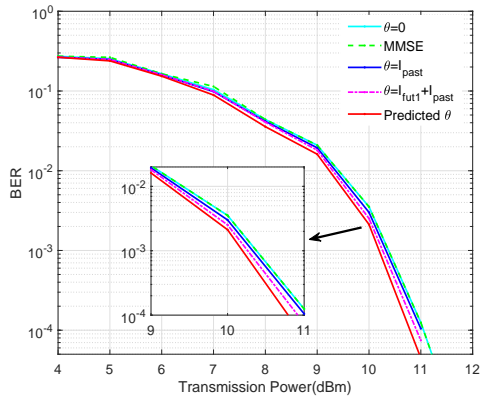}
  \caption{The experimental BERs under single path. Red solid line is the BER curve using the threshold proposed in this paper, magenta dash-dotted line is the BER curve using the threshold in \cite{Ren2020performance}, blue solid line represents the BER curve using the threshold in \cite{yao2017chaos,yao2019experimental}, green dashed line is the BER curve using zero threshold with MMSE equalization and the cyan solid line is the BER curve using zero threshold.}
\end{figure}

It can be seen from Fig. 11 that the performance of the proposed method using the decoding threshold predicted by ESN is the best although little improvement is achieved, which is consistent with the simulation result in the single path case.

\subsection{Multipath Channel Test}
To test the BER performance in outdoor multipath channel cases, we did the experiment at university campus.

\subsubsection{The First Scenario Test}
Firstly, one scenario among the classroom buildings was selected, the photo of the scenario is shown in Fig. 12(a) and the corresponding normalized estimated channel parameters are shown in Fig. 12(b).
%\begin{figure}[ht]
%  \centering
%  % Requires \usepackage{graphicx}
%  \includegraphics[width=3.5in,height=2.2in]{Fig9.eps}
%  \caption{ The photo of the test in the first scenario}
%\end{figure}
%The normalized estimated channel parameters are shown in Fig. 10, the unit of the x-axis is the delay resolution.
%\begin{figure}[ht]
%  \centering
%  % Requires \usepackage{graphicx}
%  \includegraphics[width=3.3in,height=2.2in]{Fig10.eps}
%  \caption{The normalized estimated channel parameters in the first scenario}
%\end{figure}

\begin{figure*}[ht]
\centering
\subfloat[]{\includegraphics[width=2.6in,height=1.6in]{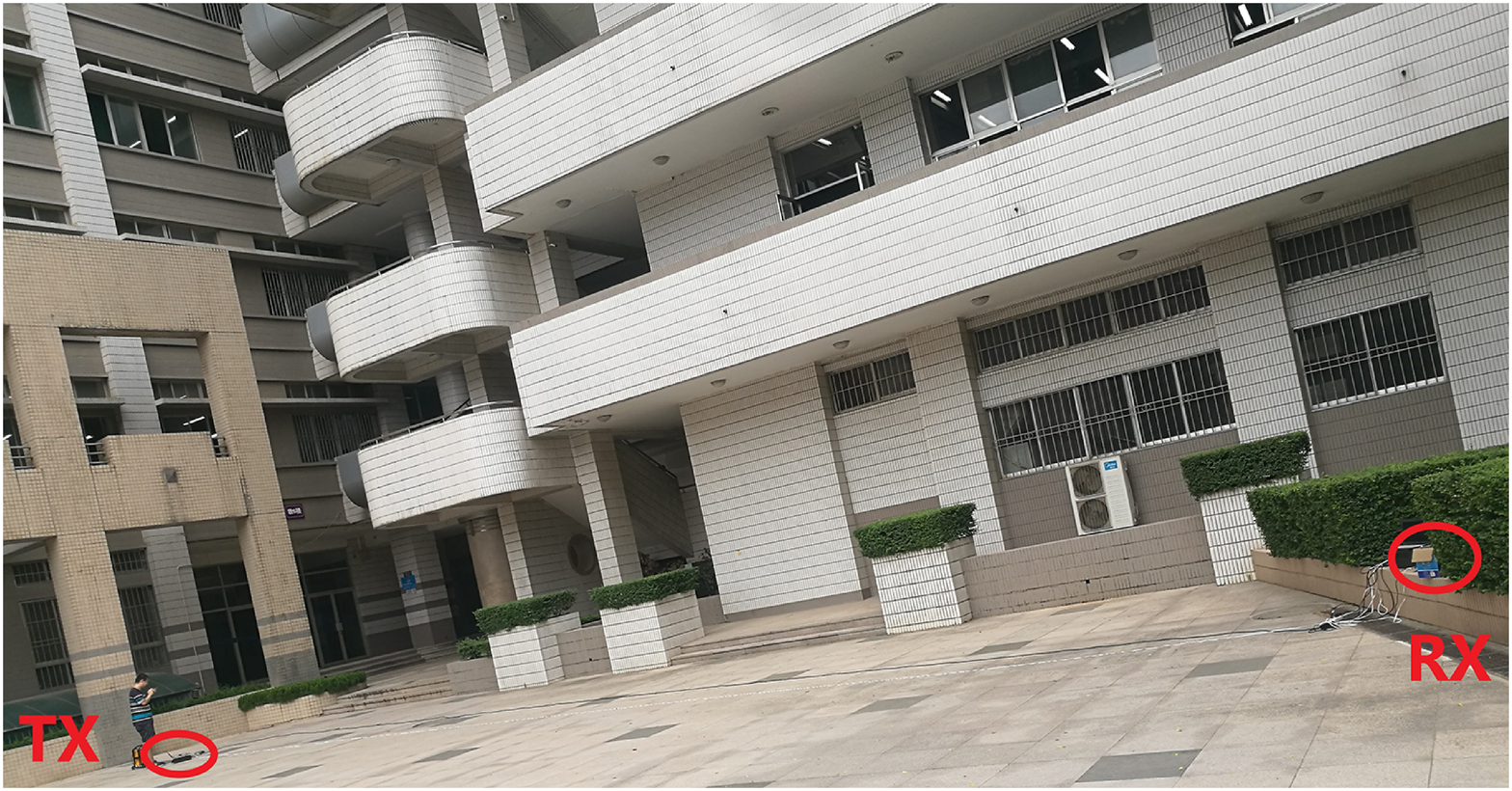}%
\label{01}}
\hfil
\subfloat[]{\includegraphics[width=2.4in,height=1.6in]{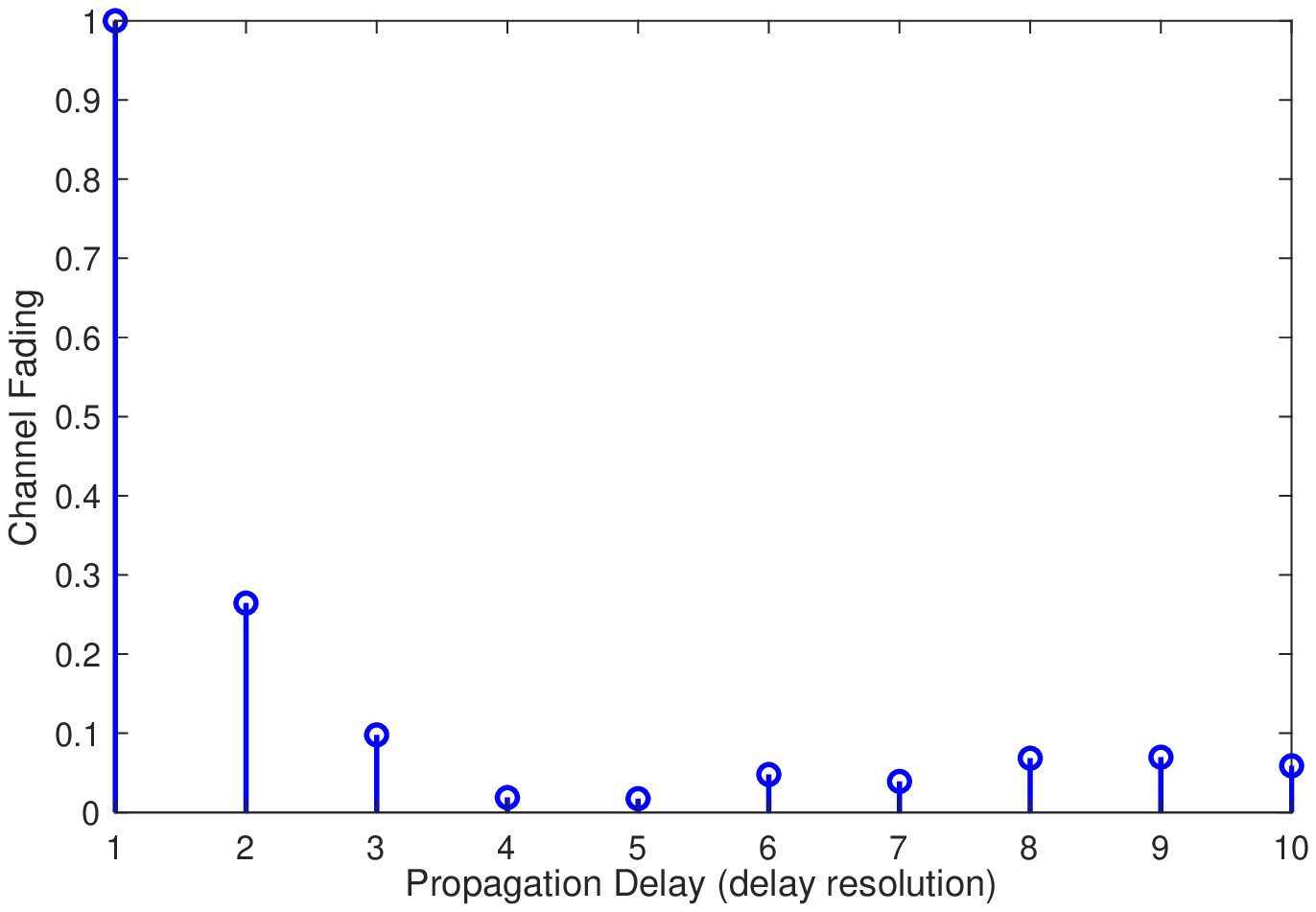}%
\label{02}}
\caption{(a) The photo of the first test scenario, (b) the normalized channel parameters estimated in the first scenario.}
\label{03}
\end{figure*}

From Fig. 12(b), we can see that there are two main paths with channel fading 1, 0.2644, the corresponding delay of two paths are 0$\mu s$ and 0.25$\mu s$, respectively.

The BER results using the proposed method in this paper, the method in \cite{Ren2020performance}, the method in \cite{yao2017chaos,yao2019experimental}, the conventional method with MMSE equalization, and the method using $\theta  = 0$ are shown in Fig. 13.
\begin{figure}[ht]
  \centering
  \includegraphics[width=3.0in,height=2.2in]{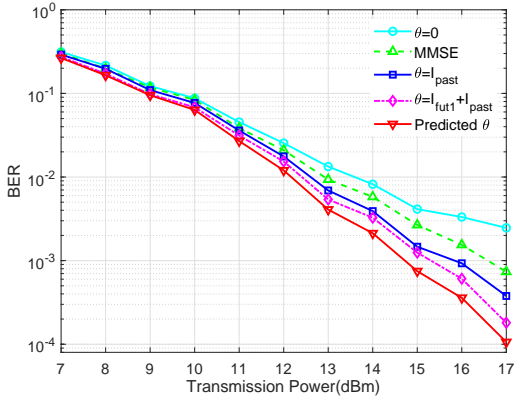}
  \caption{The experimental BER curves of different methods in the first scenario, red solid line with lower triangular markers is the BER curve using the predicted threshold in this paper, magenta dash-dotted line with diamond markers is the BER curve using the threshold in \cite{Ren2020performance}, blue solid line with square markers represents the BER curve using the threshold in \cite{yao2017chaos,yao2019experimental}, green dashed line with upper triangular markers is the BER curve using zero threshold with MMSE equalization and the cyan solid line with circle markers is the BER curve using zero threshold.}
\end{figure}

\subsubsection{The Second Scenario Test}
The second scenario test is carried out at the play yard of the campus, the photo of the second test scenario is shown in Fig. 14(a) and the corresponding normalized estimated channel parameters are shown in Fig. 14(b), the distance between TX and RX is about 30 meters.
%\begin{figure}[ht]
%  \centering
%  % Requires \usepackage{graphicx}
%  \includegraphics[width=3.5in,height=2.2in]{Fig12.eps}
%  \caption{The photo of the test in the second scenario.}
%\end{figure}
%
%The normalized estimated channel parameters are shown in Fig. 13.
%\begin{figure}[ht]
%  \centering
%  % Requires \usepackage{graphicx}
%  \includegraphics[width=3.3in,height=2.2in]{Fig13.eps}
%  \caption{The normalized estimated channel parameters in the second scenario.}
%\end{figure}

\begin{figure*}[ht]
\centering
\subfloat[]{\includegraphics[width=2.6in,height=1.6in]{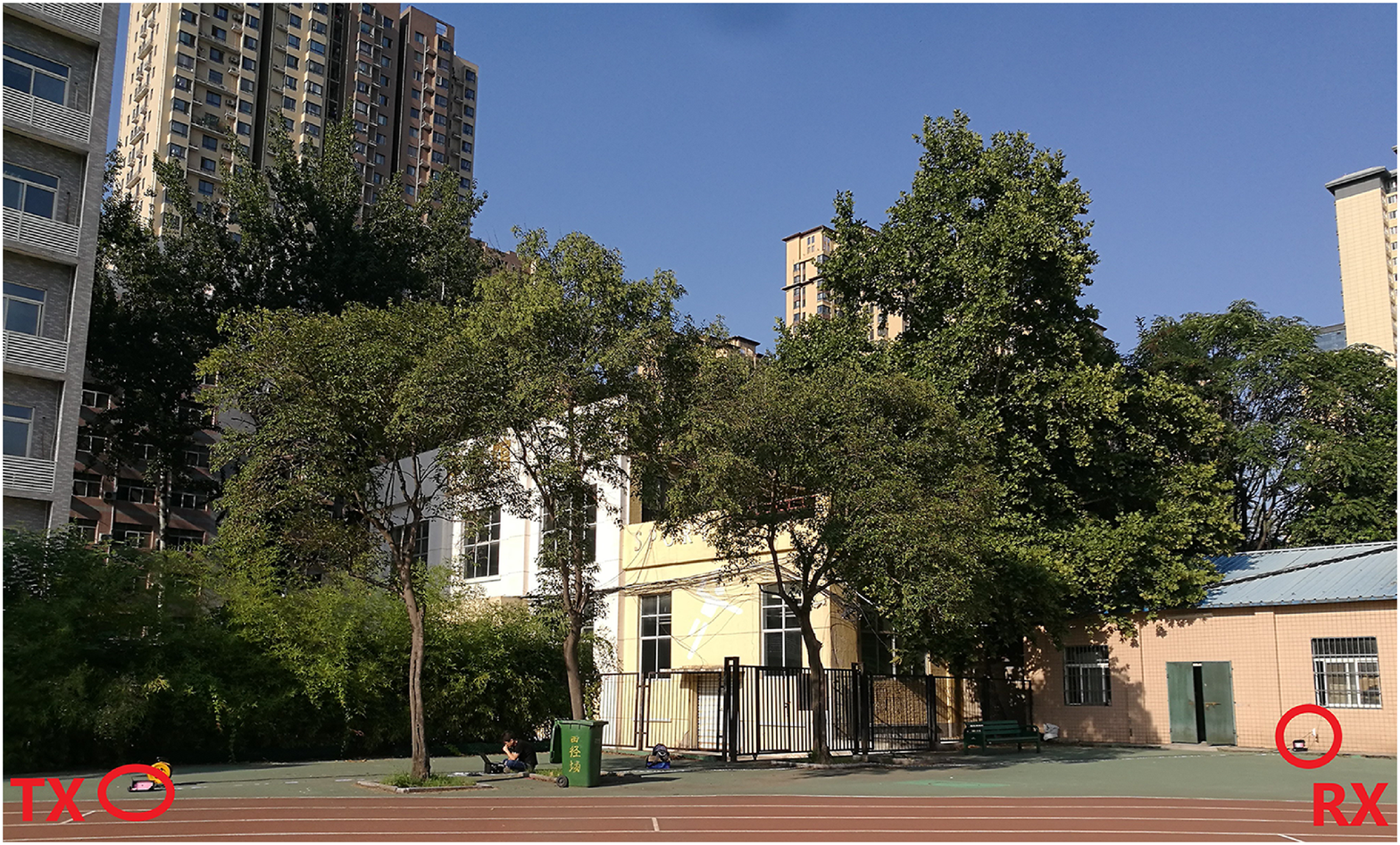}%
\label{1}}
\hfil
\subfloat[]{\includegraphics[width=2.4in,height=1.6in]{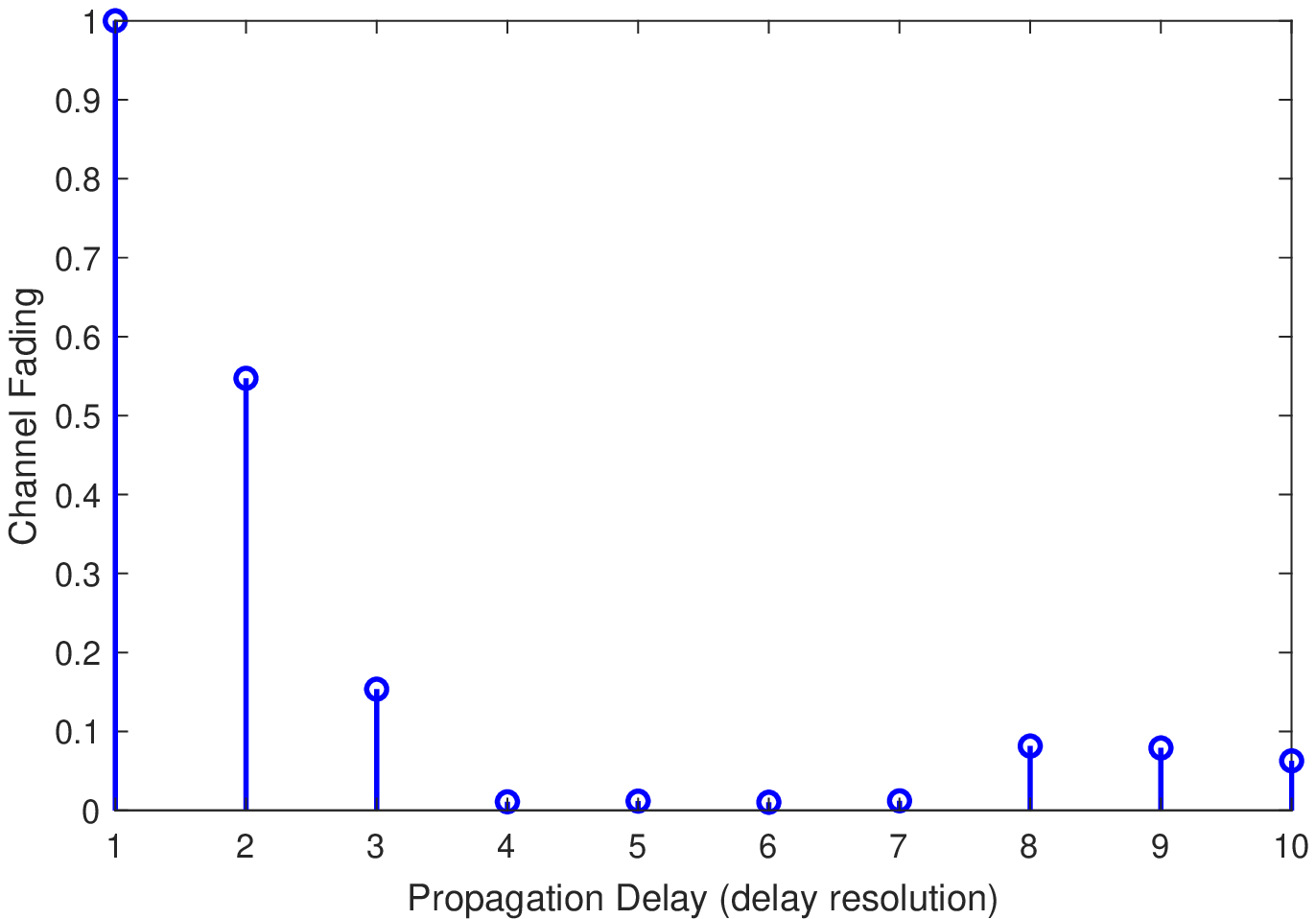}%
\label{2}}
\caption{(a) The photo of the second test scenario, (b) the channel parameters estimated in the second scenario.}
\label{3}
\end{figure*}

From Fig. 14(b), we know that there are three main paths with channel fading 1, 0.5473, 0.1535, respectively, and the corresponding time delay of three paths are 0$\mu s$, 0.25$\mu s$, 0.5$\mu s$.

The experimental BERs versus transmission power of different methods are shown in Fig. 15.

\begin{figure}[ht]
  \centering
  \includegraphics[width=2.8in,height=2.2in]{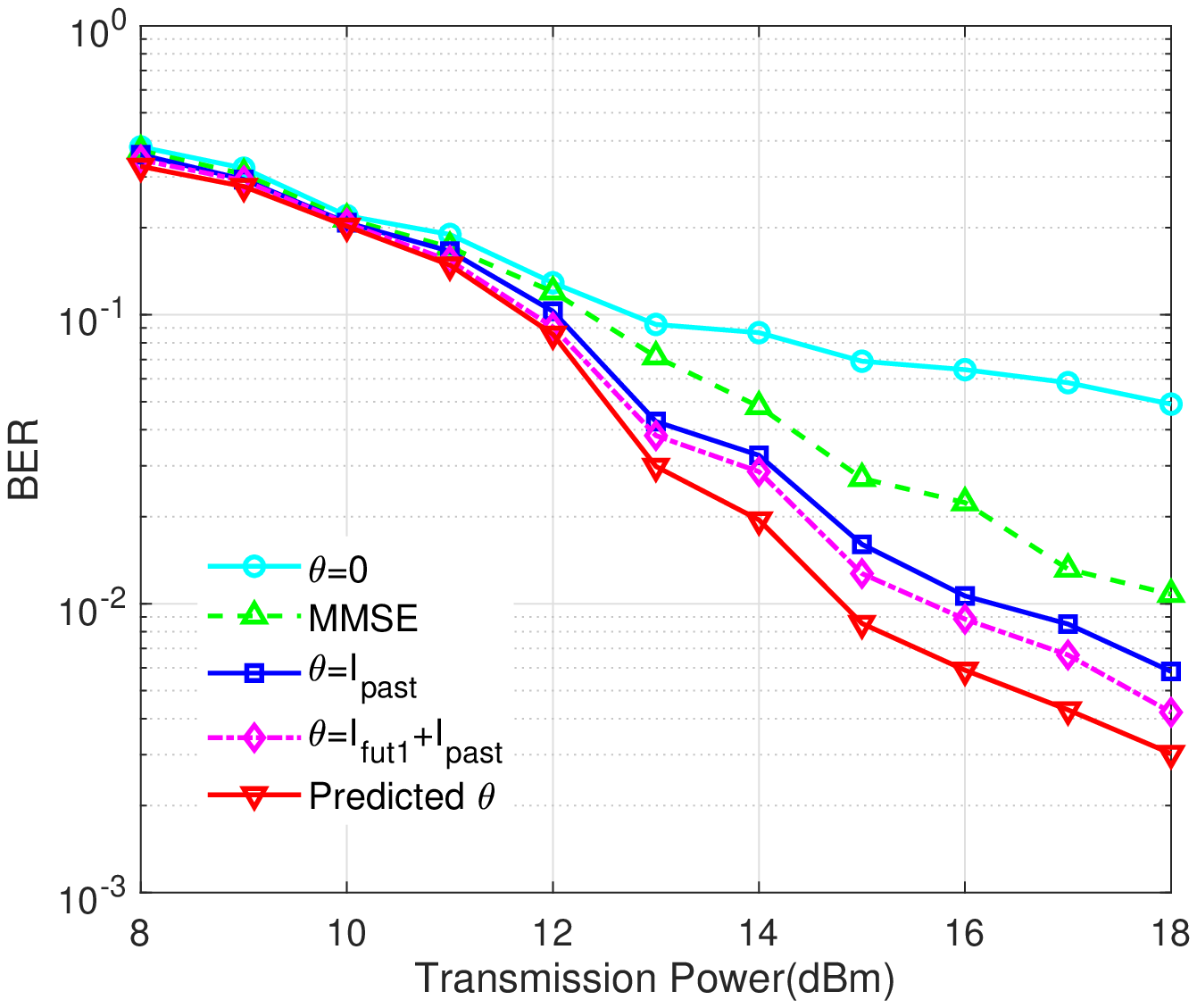}
  \caption{The experimental BERs of different methods in the second scenario, red solid line with lower triangular markers is the BER curve using the predicted threshold in this paper, magenta dash-dotted line with diamond markers is the BER curve using the threshold in \cite{Ren2020performance}, blue solid line with square markers represents the BER curve using the threshold in \cite{yao2017chaos,yao2019experimental}, green dashed line with upper triangular markers is the BER curve using zero threshold with MMSE equalization and the cyan solid line with circle markers is the BER curve using zero threshold.}
\end{figure}

In Figs. 13 and 15, the horizontal axis represents transmission power of the transmitter, which is variant to animate the signal to noise ratio variation. From Fig. 13 and Fig. 15, we can see that the traditional method using threshold $\theta  = 0$ has the worst BER, by contrast, the conventional method with MMSE equalization, the method using $\theta  = {I_{past}}$ in \cite{yao2017chaos,yao2019experimental} and the method using $\theta  = {I_{fut1}} + {I_{past}}$ in \cite{Ren2020performance} have better performance, the proposed method in this paper has the best performance. The experimental results are consistent with the simulation results, and prove the effectiveness and superiority of the proposed method.

Discussion: The proposed ESN-based threshold prediction scheme is composed of two parts, ${{\bf{W}}_{{\bf{out}}}}$ training and the optimal threshold prediction. There is no threshold calculation process as compared to the method in \cite{Ren2020performance}. Moreover, in the prediction stage, the identified channel information is not used, which reduced the additional error caused by the channel parameters estimation.

\section{Conclusion}
In this paper, an ESN-based threshold prediction method is proposed to relieve the ISI effect in CBWCS and decrease BER. The BER performance using the decoding threshold proposed in this paper is compared with the existing methods in \cite{Ren2020performance,yao2017chaos,yao2019experimental}, and the traditional methods using $\theta  = 0$, and $\theta  = 0$ with MMSE equalization. The method in \cite{yao2017chaos,yao2019experimental} can only eliminate the ISI caused by the past symbols, because the future symbols information are unavailable at the current time, the method in \cite{Ren2020performance} eliminated the ISI caused by the past symbols and one future symbol predicted by ESN. In order to further relieve the ISI caused by more future symbols and simplify the symbol decoding process, we propose ESN to predict the optimal decoding threshold directly in this paper. Both simulation and experimental results demonstrate that the ISI in CBWCS can be reduced effectively by the ESN-based predicted decoding threshold.

 %if have a single appendix:
\appendix[Computational Complexity]
% or
In this section, we compare the computational complexity of the proposed method in this paper with the method proposed in \cite{Ren2020performance}, which can be evaluated by the total number of floating-point operations (FLOPs).

The computational complexity of one threshold prediction in our method is composed of two parts, i.e., the ${{\bf{W}}_{{\bf{out}}}}$ training and the threshold prediction. In the training stage,
the main operations are the reservoir state update and ${{\bf{W}}_{{\bf{out}}}}$ calculation. In Eq. (4), calculating ${{\bf{W}}_{{\bf{in}}}}{\bf{u}}\left( {n} \right)$ requires $NK$ multiplications and $N\left( {K - 1} \right)$ summations, where $N$ is the number of neurons in the reservoir, $K$ is the number of input neurons; ${\bf{Wr}}\left( {n-1} \right)$ requires ${N^2}$ multiplications and $N\left( {N - 1} \right)$ summations; ${{\bf{W}}_{{\bf{fb}}}} \theta \left( {n-1} \right)$ requires $N$ multiplications, another two summations are needed to complete the state update; the implementation of tanh function based on piecewise approximation needs $N$ multiplications and $N$ summations \cite{ZamanlooyEfficient}. Totally, $\left( {{N^2} + NK + 2N} \right){n_{\max }}$ multiplications and $\left( {{N^2} + N\left( {K - 1} \right) + 2} \right){n_{\max }}$ summations are needed, where ${n_{\max }}$ is the number of the training bits. ${{\bf{W}}_{{\bf{out}}}}$ calculation in Eq. (5) requires ${\left( {N + K} \right)^2}\left( {{n_{\max }} - {n_0} + 1} \right) + \left( {N + K} \right)\left( {{n_{\max }} - {n_0}} \right) + {\left( {N + K} \right)^3}$ multiplications and ${\left( {N + K} \right)^2}\left( {{n_{\max }} - {n_0} - 1} \right) + \left( {N + K} \right)\left( {{n_{\max }} - {n_0} - 1} \right) + {\left( {N + K} \right)^3} + \left( {N + K} \right)\left( {N + K - 1} \right) + 1$ summations, where ${n_0}$ is the number of initial data. For one threshold prediction, the reservoir state update in Eq. (6) needs ${N^2} + NK + 2N$ multiplications and $N\left( {K - 1} \right) + {N^2} + 2$ summations, the threshold prediction in Eq. (7) requires $N + K$ multiplications and $N + K - 1$ summations. Totally, ${N^2} + NK + 3N + K$ multiplications and ${N^2} + NK + K + 1$ summations are needed.

By contrast, the previous work in \cite{Ren2020performance} predicted the waveform of the matched filter output, in a manner of predicting one sampling point per iteration. It had to predict eight points ahead iteratively by using the predicted sampling point in the last iteration as input for the next sampling point predicting, then used the final (eighth) point to compare with zero to decode (predict) one future symbol. Then the (predicted) future symbol together with the past symbols decoded and the estimated channel information to calculate a threshold, finally, the threshold is compared with the optimal SNR sampling point (i.e., the eighth sampling point) from the MF output waveform for the current symbol to decode the current symbol.

Therefore, expect training process, the method in \cite{Ren2020performance} has two points different from the proposed work in this paper, first, the trained ESN has to predict 8 times for the eighth sampling point used for decoding one future symbol; second, after the eighth sampling point prediction, the method in \cite{Ren2020performance} has to calculate the (more optimal) decoding threshold using Eq. (8) in \cite{Ren2020performance} for the current symbol decoding. By contrast, the proposed method directly predicts (one time to get) the threshold for current symbol decoding. Thus, the proposed method is intuitively efficient than the previous method in \cite{Ren2020performance}. Now, we analyze the computational complexity of the previous method in details as follows. Since the number of training data is ${n_{\max }}$, the over-sampling rate is ${N_s}$, the corresponding training sampling points are ${n_{\max }}{N_s}$, totally, $\left( {{N^2} + NK + N} \right){n_{\max }}{N_s}$ multiplications and $\left( {{N^2} + NK - N + 1} \right){n_{\max }}{N_s}$ summations are needed for the reservoir state update; ${{\bf{W}}_{{\bf{out}}}}$ calculation requires ${\left( {N + K} \right)^2}\left( {{n_{\max }}{N_s} - {n_0}{N_s} + 1} \right) + \left( {N + K} \right)\left( {{n_{\max }}{N_s} - {n_0}{N_s}} \right) + {\left( {N + K} \right)^3}$ multiplications and ${\left( {N + K} \right)^2}\left( {{n_{\max }}{N_s} - {n_0}{N_s} - 1} \right) + \left( {N + K} \right)\left( {{n_{\max }}{N_s} - {n_0}{N_s} - 1} \right) + {\left( {N + K} \right)^3} + \left( {N + K} \right)\left( {N + K - 1} \right) + 1$ summations; in the prediction, it needs the prediction operation 8 times for one future bit, i.e., $8\left( {{N^2} + NK + 2N + K} \right)$ multiplications and $8\left( {{N^2} + NK + K} \right)$ summations are needed; Here, $N$ and $K$ are defined as these for the proposed method. The additional threshold calculation requires $\left( {m + 1} \right)L$ multiplications and $mL$ summations, where $m$ represents the number of the past symbols used in the threshold calculation and $L$ is the number of the multipath.

One thing should be pointed out is that the data transmitted frame by frame, the frame is described in the respected papers. For both methods, the training process using the training data (frame pilot symbols) might use different symbols, the data symbols might be different. Here, we use the different settings according to that in the corresponding paper for calculating the average computation cost for one symbol.

To sum up, the computation cost comparison for one frame is shown in table III, the previous method in \cite{Ren2020performance} is referred to as method 1 for simplicity, and the proposed method in this paper is referred to as method 2. The parameter $K = 1$ in method 1 and $K = 16$ in method 2, and ${N_d}$ is assumed to be the number of data symbols in one frame.
 \begin{table*}[ht]\tiny
\renewcommand{\arraystretch}{2.0}

 %if using array.sty, it might be a good idea to tweak the value of
% \extrarowheight as needed to properly center the text within the cells
\caption{The computation cost comparison of two methods}
\label{table3}
\centering
% Some packages, such as MDW tools, offer better commands for making tables
% than the plain LaTeX2e tabular which is used here.
\begin{tabular}{|p{1cm}|p{3.8cm}|p{3.7cm}|p{3.7cm}|p{3.7cm}|p{3.7cm}|}
\hline\hline
\multirow{2}{*}{Cost} & \multicolumn{2}{c|}{Method 1} & \multicolumn{2}{c|}{Method 2}\\
\cline{2-5}
~& \makecell[c]{Products} & \makecell[c]{Summations} & \makecell[c]{Products} & \makecell[c]{Summations} \\
\hline
Training & $\begin{array}{l}
\left( {{N^2} + 2N} \right){n_{\max }}{N_s}\\
 + \left( {N + 1} \right)\left( {N + 2} \right)\left( {{n_{\max }} - {n_0}} \right){N_s}\\
 + {\left( {N + 1} \right)^2}\left( {N + 2} \right)
\end{array}$ & $\begin{array}{l}
\left( {{N^2} + {\rm{1}}} \right){n_{\max }}{N_s}\\
 + \left( {N + 1} \right)\left( {N + 2} \right)\left( {{n_{\max }} - {n_0}} \right){N_s}\\
 + N{\left( {N + 1} \right)^2} + {N^2}
\end{array}$ & $\begin{array}{l}
\left( {{N^2} + 18N} \right){n_{\max }}\\
 + \left( {N + 16} \right)\left( {N + 17} \right)\left( {{n_{\max }} - {n_0}} \right)\\
 + {\left( {N + 16} \right)^2}\left( {N + 17} \right)
\end{array}$ & $\begin{array}{l}
\left( {{N^2} + 15N + 2} \right){n_{\max }}\\
 + \left( {N + 16} \right)\left( {N + 17} \right)\left( {{n_{\max }} - {n_0}} \right)\\
 + {\left( {N + 16} \right)^2}\left( {N + 18} \right) + 1
\end{array}$ \\
\hline
Prediction & \makecell[c]{$8\left( {{N^2} + 3N + 1} \right)$} & \makecell[c]{$8\left( {{N^2} + N + 1} \right)$} & \makecell[c]{${N^2} + 19N + 16$} & \makecell[c]{${N^2} + 16N + 17$} \\
\hline
\tabincell{c}{Threshold\\calculation} & \makecell[c]{
$\left( {m + 1} \right)L$} & \makecell[c]{$mL$} & \makecell[c]{0} & \makecell[c]{0}\\
\hline
\tabincell{c}{Total\\operations for\\one frame} & $\begin{array}{l}
\left( {{N^2} + 2N} \right){n_{\max }}{N_s}\\
 + \left( {N + 1} \right)\left( {N + 2} \right)\left( {{n_{\max }} - {n_0}} \right){N_s}\\
 + 8{N_d}\left( {{N^2} + 3N + 1} \right)\\
 + {N_d}\left( {m + 1} \right)L + {\left( {N + 1} \right)^2}\left( {N + 2} \right)
\end{array}$ & $\begin{array}{l}
\left( {{N^2} + {\rm{1}}} \right){n_{\max }}{N_s}\\
 + \left( {N + 1} \right)\left( {N + 2} \right)\left( {{n_{\max }} - {n_0}} \right){N_s}\\
 + 8{N_d}\left( {{N^2} + N + 1} \right) + {N_d}mL\\
 + N{\left( {N + 1} \right)^2} + {N^2}
\end{array}$ & $\begin{array}{l}
\left( {{N^2} + 18N} \right){n_{\max }}\\
 + \left( {N + 16} \right)\left( {N + 17} \right)\left( {{n_{\max }} - {n_0}} \right)\\
 + {N_d}\left( {{N^2} + 19N + 16} \right)\\
 + {\left( {N + 16} \right)^2}\left( {N + 17} \right)
\end{array}$ & $\begin{array}{l}
\left( {{N^2} + 15N + 2} \right){n_{\max }}\\
 + \left( {N + 16} \right)\left( {N + 17} \right)\left( {{n_{\max }} - {n_0}} \right)\\
 + {N_d}\left( {{N^2} + 16N + 17} \right)\\
 + {\left( {N + 16} \right)^2}\left( {N + 18} \right) + 1
\end{array}$ \\
\hline
\end{tabular}
\end{table*}

According to the parameters setting in the corresponding papers, $N = 80$, ${n_{\max }} = 484$, ${n_0} = 100$, ${N_d} = 540$, $m = 4$ in method 1, $N = 112$, ${n_{\max }} = 996$, ${n_0} = 100$, ${N_d} = 1152$ in method 2, and the other parameters are the same in both methods, where ${N_s} = 16$. Thus, the computation cost comparison result using the corresponding parameters is given by table IV.
\begin{table*}[ht]
\renewcommand{\arraystretch}{1.5}
 %if using array.sty, it might be a good idea to tweak the value of
% \extrarowheight as needed to properly center the text within the cells
\caption{The computation cost comparison result of two methods}
\label{table4}
\centering
% Some packages, such as MDW tools, offer better commands for making tables
% than the plain LaTeX2e tabular which is used here.
\begin{tabular}{|c|c|c|c|c|c|}
\hline\hline
\multirow{2}{*}{Cost} & \multicolumn{2}{c|}{Method 1} & \multicolumn{2}{c|}{Method 2}\\
\cline{2-5}
~& Products & Summations & Products & Summations \\
\hline
\tabincell{c}{Total \\operations for\\one frame}& $\begin{array}{l}
104960{n_{\max }} + \\
106272\left( {{n_{\max }} - {n_0}} \right)\\
 + \left( {53128 + 5L} \right){N_d}\\
 + 538002
\end{array}$ & $\begin{array}{l}
102416{n_{\max }} + \\
106272\left( {{n_{\max }} - {n_0}} \right)\\
 + \left( {51848 + 4L} \right){N_d}\\
 + 531280
\end{array}$ & $\begin{array}{l}
14560{n_{\max }} + \\
16512\left( {{n_{\max }} - {n_0}} \right)\\
 + 14688{N_d}\\
 + 2113536
\end{array}$ & $\begin{array}{l}
14226{n_{\max }} + \\
16512\left( {{n_{\max }} - {n_0}} \right)\\
 + 14353{N_d}\\
 + 2129921
\end{array}$ \\
\hline
\tabincell{c}{Average cost\\ per symbol}& 223780.759 & 220206.133 & 41953.667 & 41344.119\\
\hline
\end{tabular}
\end{table*}

From tables III and IV, we can see that the computational complexity of the ESN method is mainly related to the training data and the number of the neurons in the reservoir. The computation cost per symbol of the proposed method is much lower than that of method 1 in \cite{Ren2020performance}. It is worth noting that the training process is just performed once in a frame, which means if the frame is longer for slow time-varying channel situation, the computational complexity is decreased as well.

To this end, we know that the ESN proposed in this method is effective and more efficient as compared to the previous method in \cite{Ren2020performance}.
%\section*


%\appendices
%% you can choose not to have a title for an appendix
%% if you want by leaving the argument blank
%\section{}
%% use section* for acknowledgment
%\section*{Acknowledgment}


\ifCLASSOPTIONcaptionsoff
  \newpage
\fi



\bibliographystyle{IEEEtran}
\bibliography{IEEEabrv,ref02}
\end{document}